# Diffusion Boundary Condition at Surface Steps

Xiaobin Niu and Hanchen Huang*

Department of Mechanical Engineering, University of Connecticut, Storrs, CT 06269, USA

**Abstract:** This Communication reports a geometrical factor that is necessary in the diffusion boundary condition across surface steps. Specifically, this factor relates adatom concentration to its spatial gradient at a surface step, and it describes the fraction of jump attempts that cross the step. In this Communication, the authors show that the factor is $1/\pi$ using theoretical formulation and further verify the formulation using numerical simulations for triangular, square, and hexagonal surface lattices.

* To whom correspondence should be addressed; E-mail: hanchen@uconn.edu

## 1. Introduction

Diffusion boundary conditions are critically important in theoretical studies of crystal growth, during which atomic islands form and evolve. On top of the atomic islands, adatoms may also diffuse across surface steps. For many crystals, there is an extra energy barrier for adatoms to diffuse across a step downward. This extra barrier is the Ehrlich-Schwoebel barrier [1,2], and it depends on the thickness of the step [3-7]. The extra barrier defines whether epitaxy will be layer-by-layer or three-dimensional growth, and how large the grain size will be in a polycrystalline thin film [8-10]. Numerous theoretical studies – in addition to computer simulations and experiments – have examined crystal growth using this diffusion process as a boundary condition [9-12].

Within the framework of theoretical studies, adatoms arrive at the top of an island at a rate of $F$, and the evolution equation of adatom concentration $C$, as a function of time $t$ and position $r$, is: $\partial C(r,t)/\partial t = D\nabla^2 C(r,t) + F$ before the nucleation of new clusters; here, $D$ is the adatom diffusion coefficient. To solve this equation, even under quasi-steady state conditions in an effectively circular island area, two boundary conditions are necessary. One of the boundary conditions is: $-DdC/dr = \alpha \upsilon \exp(-\Delta E/kT)C$ at the step that bounds the island. Here, $\Delta E$ is the Ehrlich-Schwoebel barrier, $kT$ is the product of Boltzmann constant and temperature, and $\upsilon$ is the diffusion jump frequency of an adatom on the island. In particular, $\alpha$ is a geometrical factor that defines the fraction of diffusion jumps that are directed towards the outside of the island. However, the value of $\alpha$ remains undecided [9,10] or is effectively used as $1/4$ on two-dimensional surfaces [11,12].

This Communication uses theoretical formulation to show that $\alpha$ is $1/\pi$, and uses numerical simulations to verify the formulation. In particular, for triangular surface lattice, the simulations give $\alpha$ as 0.33, close to $1/\pi$. When the coordination of surface lattice decreases, the $\alpha$ value becomes slightly larger than $1/\pi$.

## 2. Theory

We consider a circular island of radius $R$, part of which is shown in Figure 1. An adatom at $O'$ may diffuse outside the island within one atomic jump of distance $a$. With the fixed jump distance $a$, the adatom can land anywhere on the perimeter of the circle centered around $O'$, but only the jumps onto the arc $\stackrel{\frown}{AC'B} = l$ result in diffusion outside of the island. Therefore the fraction of jumps that direct to outside of the island, the factor $\alpha$ for a particular adatom starting at $O'$, is given by $l/2\pi a$. Since the adatom location $O'$ can be anywhere within the circular shell of thickness $a$ inside the island boundary, the factor $\alpha$ is the geometrical average over the shell area:

$$\alpha = \int_{R-a}^{R} (l/2\pi a) 2\pi r dr \Big/ \int_{R-a}^{R} 2\pi r dr = \int_{R-a}^{R} (l/a) r dr \Big/ \int_{R-a}^{R} 2\pi r dr. \tag{1}$$

When the radius $R \gg a$, the distance between an adatom position and the center of the island is:

$$r \approx R - a \cos\frac{\theta}{2}, \tag{2}$$

$$dr \approx \frac{1}{2} a \sin\frac{\theta}{2}, \tag{3}$$

The integration of $r$ from $R - a$ to $R$ is equivalent to the integration of $\theta$ from 0 to $\pi$:

$$\int_{R-a}^{R} (l/a) r dr = \int_{0}^{\pi} \left( \frac{1}{2} Ra\theta \sin\frac{\theta}{2} - \frac{1}{4} a^2 \theta \sin\theta \right) d\theta = 2Ra - \frac{1}{4}\pi a^2, \tag{4}$$

$$\int_{R-a}^{R} 2\pi r dr = \int_{0}^{\pi} \left( \pi Ra \sin\frac{\theta}{2} - \frac{1}{2}\pi a^2 \sin\theta \right) d\theta = 2\pi Ra - \pi a^2. \tag{5}$$

Substitution of these expressions into equation (1) gives:

$$\alpha = \int_{R-a}^{R} (l/a) r dr \Big/ \int_{R-a}^{R} 2\pi r dr = \frac{2Ra - \pi a^2/4}{2\pi Ra - \pi a^2} \xrightarrow{R \to \infty} \frac{1}{\pi}. \tag{6}$$

As a brief discussion and generalization, we note that adatoms may also diffuse into an island, or into the circle of Figure 1 from outside. Following the same formulation, we can derive that the factor $\alpha$ is:

$$\alpha = \int_{R+2a}^{R+a} (l/a) r dr \Big/ \int_{R+2a}^{R+a} 2\pi r dr = \frac{2(R+a)a + \pi a^2/4}{2\pi (R+a)a + \pi a^2} \xrightarrow{R \to \infty} \frac{1}{\pi}. \tag{7}$$

For sufficiently large islands, $R \gg a$, the factor $\alpha$ is $1/\pi$ for adatom diffusion both into and out of an island.

It is important to note that we assume $R \gg a$ in the formulation with the following justification. This assumption effectively means an average over diffusion processes across all kinds of surface steps, and is valid when surface islands are bounded by non-straight steps. In terms of growth conditions, either insufficient diffusion or high temperature roughening will ensure the validity. As the steps become straight, the fraction can be determined based on crystal symmetry; for example, the factor $\alpha$ should be 1/3 for a <110> step on face-centered-cubic {111} surface.

## 3. Numerical Verification

To verify the theoretical formulation, we numerically determine the factor $\alpha$ for three surface lattice structures: (1) triangular lattice corresponding to a face-centered-cubic {111} surface, (2) square lattice corresponding to a face-centered-cubic {100} surface, and (3) hexagonal lattice corresponding to agraphitic {0001} surface. In each simulation, we populate a circular area of radius $R$ by lattice sites, and place an adatom at a lattice site within diffusion jump distance $a$ inside the edge of the circle. Then, we move the adatom to its nearest neighboring lattice site at equal probability. The fraction of moves that land the adatom at a lattice site outside the circle of radius is $\alpha$. The simulation is repeated for different radii $R$, with increment of diffusion jump distance $a$.

As shown in Figure 2, the average of factor $\alpha$ converges to 0.33 for triangular lattice, in favorable comparison with the theoretical value of $1/\pi$ or 0.32. The difference between them results from the discrete nature of the lattice. According to the theoretical formulation, the smallest fraction of jumping outside a circle of radius $R$ is zero; because $l$ can be as small as zero. In numerical simulations, however, the smallest fraction can never be zero. Even for a triangular lattice, the diffusing adatom at the step edge on the island has at least one nearest neighbor outside of the circle, by the very definition of step edge adatom. Therefore, the smallest fraction is 1/6 instead of zero, and the average $\alpha$ from numerical simulation excludes any events with fractions smaller than 1/6, and therefore is larger than the theoretical value. As the coordination in the surface lattice goes from 6 to 4 (from triangular to square lattice), this difference increases since the smallest fraction in simulations is now 1/4; the same trend applies when the coordination goes further to 3 for hexagonal lattice.

## 4. Conclusion

In summary, this Communication reports a geometrical factor α that relates adatom concentration with its spatial gradient at surface steps. Our theoretical formulation shows that the factor $\alpha$ is $1/\pi$, and our numerical simulations further verify the formulation and reveal small variations as surface lattice structure changes. For three common surface lattice structures, the numerical simulations show that the factor $\alpha$ is 0.33 for triangular lattice, 0.35 for square lattice, and 0.38 for hexagonal lattice. In passing, we also note that the factor $\alpha$ applies to adatom diffusion both into and out of an island.

**Acknowledgment:** The authors gratefully acknowledge financial support from the National Science Foundation (CMMI-0856426), and helpful discussions with Dr. Longguang Zhou.

**Figure captions:**

Fig. 1. Schematic of an adatom at *O′* diffusing across an island of radius *R*.

Fig. 2. Geometrical factor $\alpha$ as a function of island radius *R* in the unit of diffusion jump distance *a*; for three different surface lattices.